\documentclass{llncs}

\usepackage{makeidx}  
\usepackage{color}
\usepackage{graphicx}
\usepackage[nounderscore]{syntax}
\usepackage{courier}
\usepackage{hyperref}
\usepackage{amsmath}
\usepackage{float}
\usepackage{url}
\usepackage[labelfont=bf, labelsep=period]{caption}

\newfloat{requirement}{h}{req}
\floatname{requirement}{Req.}

\begin{document}
%
%
\pagestyle{plain}  
%

\title{Requirements Analysis of a Quad-Redundant Flight Control System}
\author{John Backes\inst{1}, Darren Cofer\inst{1}, Steven Miller\inst{1}, Michael W. Whalen\inst{2}}
\institute{Rockwell Collins, Bloomington MN 55438 \and University of Minnesota, Minneapolis MN 55455}
\maketitle
\begin{abstract}
In this paper we detail our effort to formalize and prove requirements for the Quad-redundant Flight Control System (QFCS) within NASA's Transport Class Model (TCM). 
We use a compositional approach with assume-guarantee contracts that correspond to the requirements for  software
components embedded in an AADL system architecture model.  This approach is designed to exploit the verification effort and artifacts that
are already part of typical software verification processes in the avionics domain.  Our approach is supported by 
an AADL annex that allows specification of contracts along with a tool, called AGREE, for performing compositional verification.
The goal of this paper is to show the benefits of a compositional verification
approach applied to a realistic avionics system and to demonstrate
the effectiveness of the AGREE tool in performing this analysis.

%
\end{abstract}

\section{Introduction}

Modern aircraft are complex cyber-physical systems with safety and security
requirements that must be satisfied by their onboard software. As these systems
have grown in complexity, their verification has become the single most costly
development activity~\cite{crum04}. The verification costs of even more complex systems in the
future will impact safety, not just through an increasing incidence of errors
and unforeseen interactions, but by delaying and preventing the deployment of
crucial safety functions.

In a NASA-funded project with University of Minnesota and University of Iowa we
are addressing these challenges by developing compositional reasoning methods
that will permit the verification of systems that exceed the complexity limits
of current approaches. Our approach is based on:

\begin{itemize}
\item Modeling the system architecture using standard notations that will be
usable by systems and software engineers.

\item Developing a sophisticated translation framework that automates the translation
of these models for analysis by powerful general-purpose verification engines
such as SMT-based model checkers.

\item Developing techniques for compositional verification based on the system
architecture to divide the verification task into manageable, reusable pieces.
\end{itemize}

This approach has the potential to significantly reduce verification costs by identifying and correcting
system design errors early in the life cycle rather than waiting until system
integration. We are validating our approach and our tools on a
realistic fault-tolerant flight control system model.  The Quad-redundant Flight Control System
(QFCS) has been designed by NASA as a suitable control system for its Transport Class Model
(TCM) aircraft.

\sloppypar
Our compositional approach is designed to exploit the verification effort and artifacts that
are already part of typical software component verification processes. Each component in the system 
model is annotated with an assume/guarantee contract that includes the requirements ({\em guarantees}) and 
environmental constraints ({\em assumptions}) that were specified and verified as
part of its development process.   We then reason about the system-level behavior
based on the interaction of the component contracts. By partitioning the verification
effort into proofs about each subsystem within the
architecture, the analysis will scale to handle large system designs. Additionally, the
approach naturally supports an architecture-based notion of requirements refinement:
the properties of components necessary to prove a system-level property in effect
define the requirements for those components.

There were two objectives in using this verification approach. The first was to
reuse the verification already performed on components. The
second was to enable distributed, parallel development of components via 
{\em virtual integration}.  In this process, we specify formal component-level 
requirements, demonstrate that they are sufficient to prove system guarantees, 
and then use these requirements as specifications for suppliers.  If the suppliers' 
implementations meet these specifications, we have a great deal of confidence that 
the integrated system will work properly.


We have chosen the Architecture Analysis and Design Language (AADL) as our system
architecture modeling language \cite{aadl}.   AADL was designed for embedded, real-time,
distributed systems and so is a good fit for our domain.  It provides the constructs needed
to model embedded systems such as threads, processes, processors, buses, and memory.
It is sufficiently formal for our purposes, and is extensible through the use of language
annexes that can initiate calls to separately developed analysis tools.

We have implemented our compositional reasoning methodology in a tool called {\it AGREE:
Assume-Guarantee Reasoning Environment}.  AGREE is implemented as an Eclipse plugin
and is designed to work with the open source OSATE AADL tool developed by the Software
Engineering Institute~\cite{osate}.  AGREE is able to check the correctness of behavioral properties defined
by the composition of component contracts, check component contracts for inconsistencies,
and determine whether a component contract has any possible realization.
AGREE makes use of the AADL annex mechanism to annotate
models with contracts corresponding to formal assumptions and guarantees about their
behaviors.  AGREE is open source software and is available at \url{http://github.com/smaccm}.

The goal of this paper is to show the benefits of a compositional verification
approach applied to a realistic avionics system and its requirements, and to demonstrate
the effectiveness of the AGREE tool in performing this analysis.

\section{Compositional Verification with AGREE}\label{sec:comp-v}

In this section we briefly describe the rules that AGREE uses to create compositional proofs. A more complete description is in~\cite{CoferNFM2012} and a proof of correctness of these rules is provided in~\cite{AgreeUsersGuide,CoferNFM2012}.

AGREE is a language and a tool for compositional verification of AADL models.  The behavior of a model is described by \textit{contracts} specified on each component.  A contract contains a set of \textit{assumptions} about the component's inputs and a set of \textit{guarantees} about the component's outputs.  The assumptions and guarantees may also contain predicates that reason about how the state of a component evolves over time. The state transitions of each component in the model occur synchronously with every other component (i.e., each component runs on the same clock).  The guarantees of a component must be true provided that the component's assumptions have always been true.  The goal of the analysis is to prove that a component's contract is entailed by the contracts of its subcomponents.

Formally, let a system $S : (A,G,C)$ consist of a set of assumptions $A$, guarantees $G$, and subcomponents $C$. We use the notation $S_g$ to represent the conjunction of all guarantees of $S$ and $S_a$ to represent the conjunction of all assumptions of $S$. Each subcomponent $c \in C$ is itself a system with assumptions, guarantees, and subcomponents. The goal of our analysis is to prove that the system's guarantees hold as long as its assumptions have always held.  This is accomplished by proving that Formula~\ref{fml:obligation} is an invariant.

\begin{center}
\begin{equation}\label{fml:obligation}
\mathbf{H}(S_a) \rightarrow S_g
\end{equation}
\end{center}

The predicate $\mathbf{H}$ is true if its argument has held \textit{historically} (i.e., the expression has been true at every time step up until and including now). In order to prove that Formula~\ref{fml:obligation} is invariant, we prove that the assumptions of all the subcomponents of system $S$ hold under the assumptions of $S$. This invariant is shown in Formula~\ref{fml:strong-assump}.

\begin{center}
\begin{equation}\label{fml:strong-assump}
\bigwedge_{c\in C} \Big[ \mathbf{H}(S_a) \rightarrow c_a \Big]
\end{equation}
\end{center}

This formula is actually stronger than what we need to prove.  It may be the case that the assumptions of certain subcomponents are satisfied by the guarantees of other subcomponents (and possibly the guarantees of the component itself at previous instances in time).  This weaker invariant is shown in Formula~\ref{fml:weak-assump}.

\begin{center}
\begin{equation}\label{fml:weak-assump}
 \bigwedge_{c\in C} \Big[ \Big( \mathbf{H}(S_a) \wedge \bigwedge_{w \in C} \mathbf{Z}(\mathbf{H}(w_g))\wedge\bigwedge_{v \in C, c \neq v} \mathbf{H}(v_g) \Big) \rightarrow c_a \Big]
\end{equation}
\end{center}

The predicate $\mathbf{Z}$ is true in the first step of a trace and thereafter is true iff its argument was true in the previous time step. 

However, this formula may not be sound when the connections between components form cycles.  One could imagine a scenario where the assumptions of each of two components are true precisely because of the guarantees of the other component (i.e., $w_g \rightarrow v_a $ and $v_g \rightarrow w_a$ for $w,v \in C$ and $w \neq v$). Suppose components $w$ and $v$ both assume that their inputs are positive, and they guarantee that their outputs are positive.  If the output of $w$ is connected to the input of $v$, and $v$'s output is connected to $w$, the state of the system is improperly defined.  To avoid this problem, AGREE  creates a total ordering of a system's subcomponents.  It uses this ordering to determine which subcomponent guarantees are used to prove the assumptions of other subcomponents.  This slight modification of Formula~\ref{fml:weak-assump} is shown in Formula~\ref{fml:correct-assump}.

\begin{center}
\begin{equation}\label{fml:correct-assump}
\bigwedge_{c\in C} \Big[ \Big( \mathbf{H}(S_a) \wedge \bigwedge_{w \in C}\mathbf{Z}(\mathbf{H}(w_g))\wedge\bigwedge_{v \in C, v < c} \mathbf{H}(v_g) \Big) \rightarrow c_a \Big]
\end{equation}
\end{center}

If Formula~\ref{fml:correct-assump} is invariant then Formula~\ref{fml:obligation} is proven to be invariant by showing that the system assumptions and subcomponent guarantees satisfy the system guarantees.  Formally, if Formula~\ref{fml:correct-assump} is invariant, then Formula~\ref{fml:guar-check} implies Formula~\ref{fml:obligation}.

\begin{center}
\begin{equation}\label{fml:guar-check}
 \mathbf{H}(S_a) \wedge \bigwedge_{c \in C} \mathbf{H}(c_g) \rightarrow S_g
\end{equation}
\end{center}

AGREE uses a syntax similar to Lustre to express a contract's assumptions and guarantees~\cite{lustre}. AGREE translates an AADL model annotated with AGREE annexes into Lustre corresponding to Formulas~\ref{fml:correct-assump} and~\ref{fml:guar-check} and then queries a user selected model checker. AGREE then translates the results from the model checker back into OSATE so they can be interpreted by the user. For this project we have used both the Kind 2.0 and JKind model checkers~\cite{kind2,Jkind}.

In Section~\ref{sec:reqs} we describe some examples of guarantees that were written in AGREE to model some of the requirements in the QFCS architecture.  However, the examples are presented here in a simple first order logic syntax to make them more concise and readable.

\section{Requirements Formalization}\label{sec:reqs}
\setcounter{equation}{0} 

\begin{figure}
\begin{center}
\includegraphics[scale=0.5]{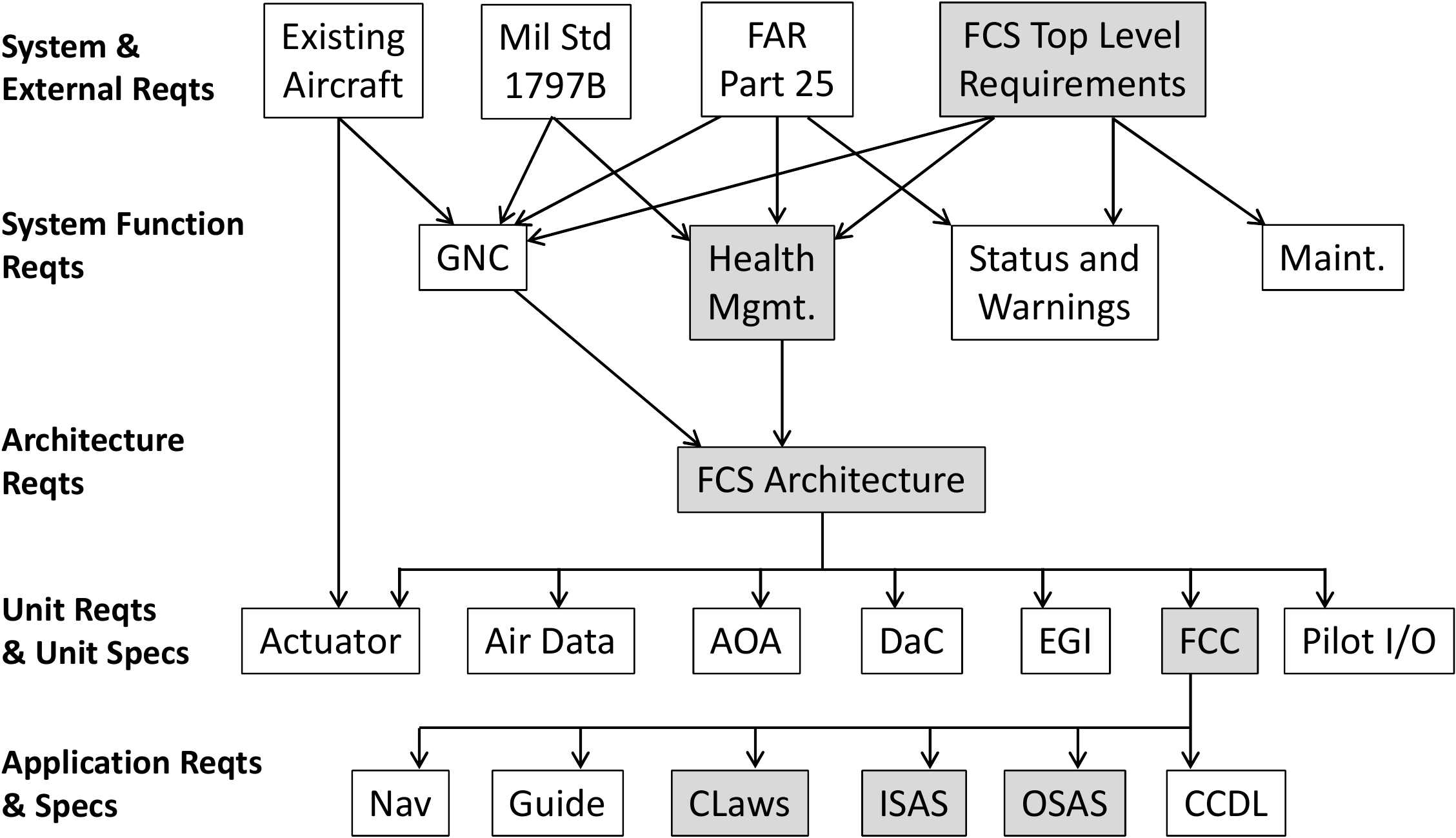}
\end{center}
\caption{The QFCS requirements hierarchy (those in grey were included in our analysis)}
\label{fig:reqs-hier}
\end{figure}

We are using NASA's TCM aircraft simulation model~\cite{hueschen2011} as a realistic example to demonstrate and validate our compositional reasoning work. 
The TCM was not originally developed with a set of requirements, but other researchers have created a set of requirements representative of those that would be necessary to certify an aircraft for operation in the national airspace system~\cite{brat2014}. These requirements were developed hierarchically with different requirements being assigned to different levels of the system architecture, all the way down to the major software components.  The requirements hierarchy is shown in Figure~\ref{fig:reqs-hier}.

\subsection{QFCS Architecture}

\begin{figure}[!h]
\begin{center}
\includegraphics[scale=0.5]{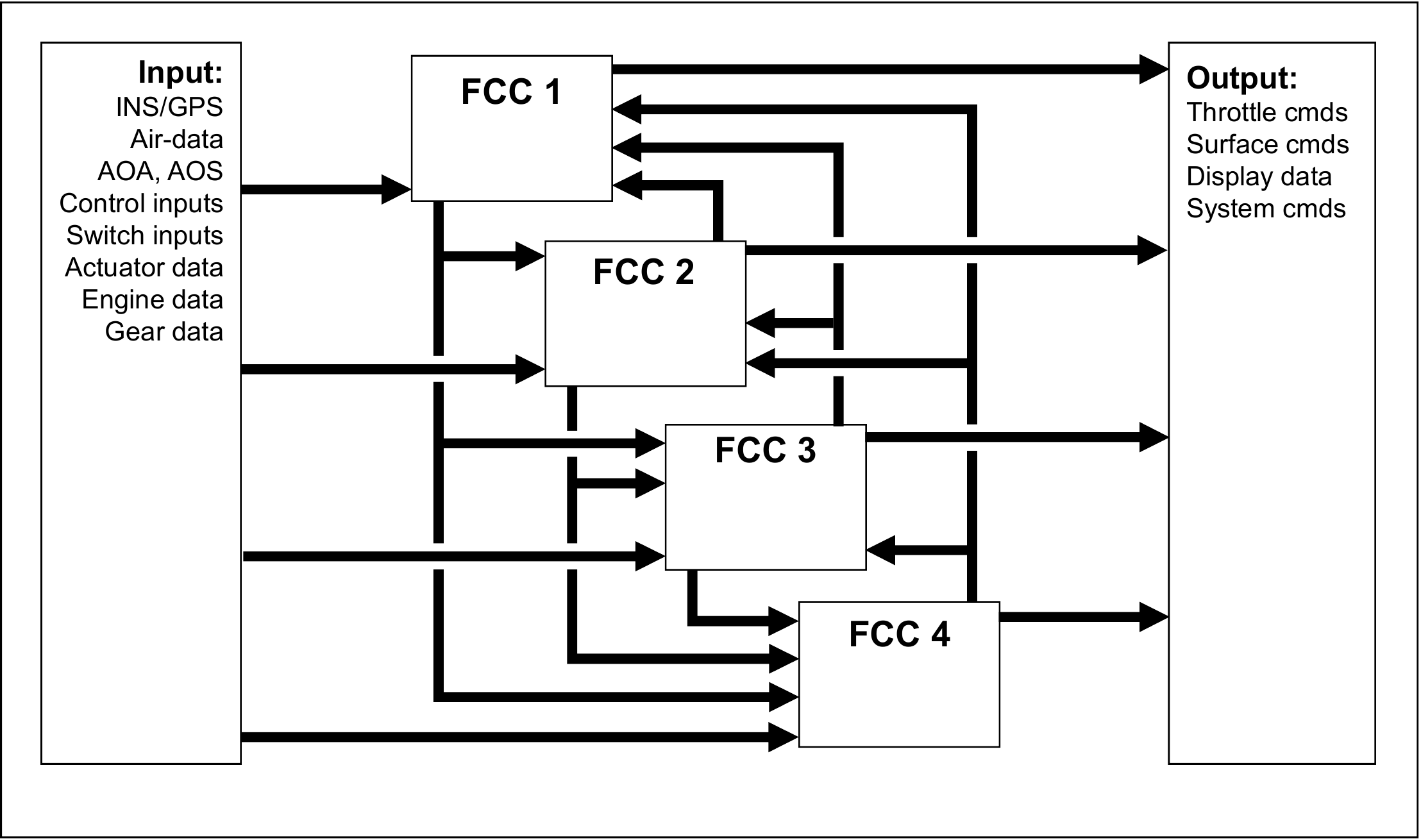}
\end{center}
\caption{The QFCS architecture with four flight control computers}
\label{fig:fcc}
\end{figure}

The QFCS is a quad-redundant flight control system for the TCM consisting of four cross-checking flight control computers (FCC), as shown in Figure~\ref{fig:fcc}.  The QFCS model was developed in Simulink\textsuperscript{\textregistered} and includes models of the aircraft's control laws, sensors, and actuators, and interacts with the TCM aerodynamics model. The fault tolerance logic was not originally part of this model, but was added to the simulation in parallel during our project.

Our work focused on formalizing requirements for five components of the QFCS hierarchy: the Flight Control System (FCS), the Flight Control Computers (FCC), the Output Signal Analysis and Selection component (OSAS), the Input Signal Analysis and Selection component (ISAS), and the Control Laws (CLAW).  The FCS consists of four individual FCCs, and each FCC includes a single OSAS, ISAS, and CLAW component, as well as several other components. We focused on formalizing the requirements for these components for a couple reasons. First, others were working to formalize some of the other components using different techniques in parallel with this work~\cite{brat2014}. Second, the requirements for these components had a much clearer path to formalization compared to the other component requirements.

\begin{figure}[!h]
\begin{center}
\includegraphics[scale=0.6]{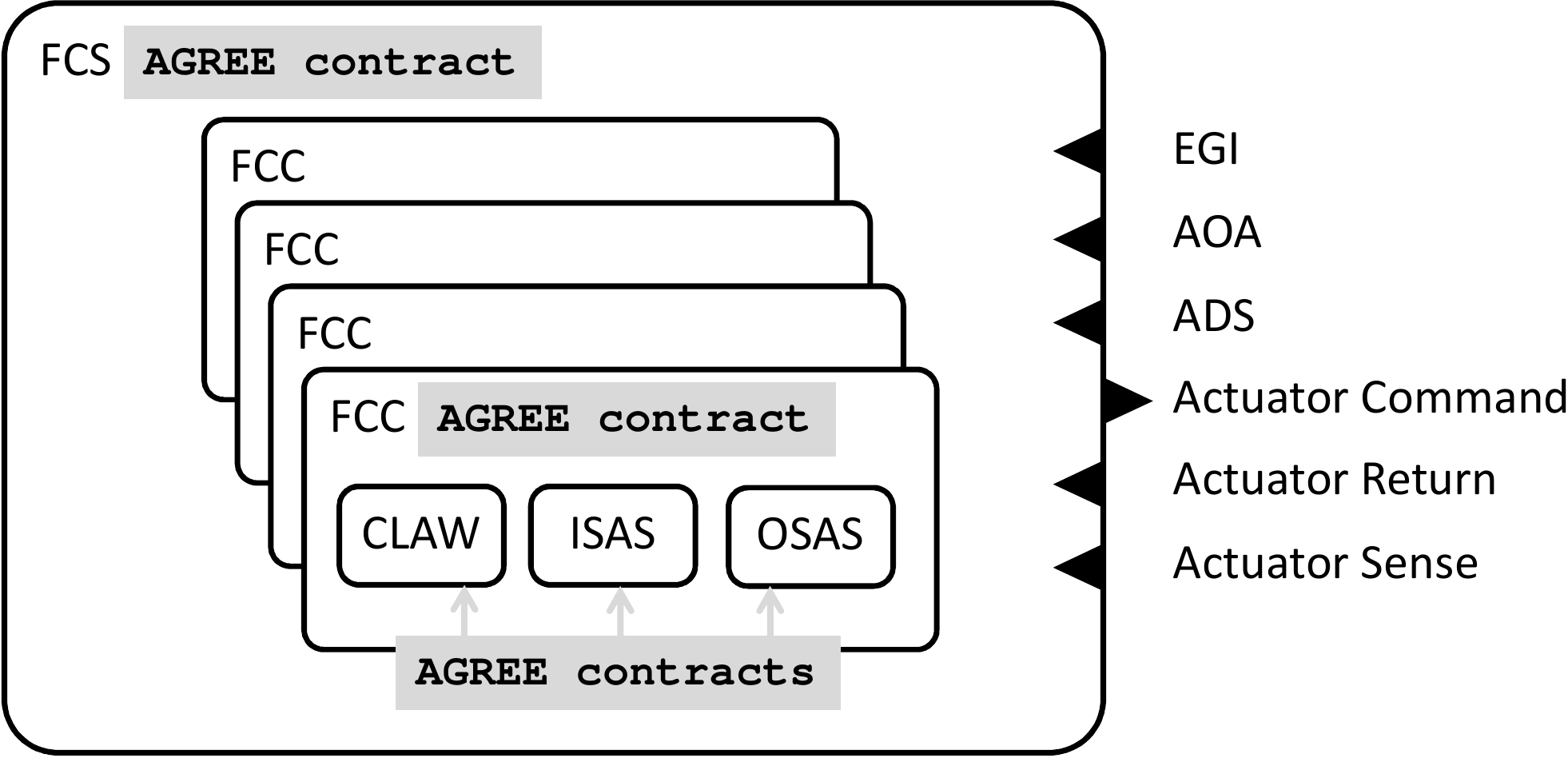}
\end{center}
\caption{The QFCS component hierarchy}
\label{fig:qfcs-diagram}
\end{figure}

The FCS component hierarchy is shown in Figure~\ref{fig:qfcs-diagram}. These components were modeled in AADL with the same interfaces and connections described in the QFCS Simulink\textsuperscript{\textregistered} model. The requirements for the QFCS were taken from the hierarchy of requirements shown in Figure~\ref{fig:reqs-hier} and were formalized and assigned as assume guarantee contracts to the relevant QFCS components in the AADL model.  AGREE was used to show that the requirements at each level of the component hierarchy were satisfied by the requirements of their direct subcomponents. Explicitly, the requirements formalized for the FCS were proven to hold by the composition of the requirements of the four FCCs. Additionally, the requirements of each FCC were satisfied by the requirements of the OSAS, ISAS, and CLAW components. This section lists examples of some of the English language requirements that were formalized for some of these components. In particular, we discuss requirements related to the actuator signals that are sent from each flight control computer.

In the remainder of this section we list examples of some of the English language requirements that we formalized for some of these components. In particular, we discuss requirements related to the actuator signals that are sent from each flight control computer.

\subsection{Flight Control System}

The FCS requirements make up the ``top level'' properties that should be satisfied by the composition of the requirements of all of the components within the FCS.  Many of the FCS requirements reference functions that we have not yet modeled, including as  Guidance Navigation and Control, Maintenance Function, and Status and Warning.  We have chosen to focus our analysis on the fault tolerance requirements for the FCS.  The top level FCS requirement for the fault handling logic is shown in Requirement~\hyperref[req:fcs]{FCS-120}.

\begin{requirement}
\begin{center}
\noindent\fbox{%
    \parbox{280pt}{%
\textbf{FCS-120} - The Health Management Function (HM) shall detect and mitigate Flight Control System faults.
}%
}\vspace{-2em}
\end{center}
\label{req:fcs}
\end{requirement}

This statement is certainly too vague to be formalized. There is no guidance given on what qualifies as ``mitigating a fault''. However, Requirement~\hyperref[req:fcs]{FCS-120} depends on many sub requirements that are more precise.  Among these are Requirements~\hyperref[req:single]{HM-220}~and~\hyperref[req:double]{HM-240}.

\begin{requirement}
\begin{center}
\noindent\fbox{%
    \parbox{280pt}{%
\textbf{HM-220} - The Health Management Function shall provide Cooper Harper Level 4 Handling Qualities after any single LRU, LRU function, or LRU IO signal failure.
}%
}
\vspace{-2em}
\end{center}
\label{req:single}
\end{requirement}

\begin{requirement}
\begin{center}
\noindent\fbox{%
    \parbox{280pt}{%
\textbf{HM-240} - The Health Management Function shall provide Cooper Harper Level 4 Handling Qualities after any dual simultaneous LRU, LRU function, or LRU IO signal failures including actuator runaways and jams not shown to be extremely improbable.
}%
}
\end{center}
\label{req:double}
\end{requirement}

Requirements~\hyperref[req:single]{HM-220}~and~\hyperref[req:double]{HM-240} are also challenging to formalize because they require that the aircraft meet a specific Cooper Harper rating~\cite{cooper69}. A Cooper Harper rating is a subjective measurement used to describe the ease with which a pilot is able to operate an aircraft. However, there are some objective properties that are related to these statements.  We propose Requirements~\hyperref[req:bound]{HM-240a}~and~\hyperref[req:mono]{HM-240b} as properties that can be stated precisely and are necessary for satisfying Requirements~\hyperref[req:single]{HM-220}~and~\hyperref[req:double]{HM-240}.

\begin{requirement}
\begin{center}
\noindent\fbox{%
    \parbox{280pt}{%
\textbf{HM-240a} - The average of the signals sent to any given actuator is bounded regardless of how many LRU failures occur.
}%
}
\end{center}
\label{req:bound}
\end{requirement}

\begin{requirement}
\begin{center}
\noindent\fbox{%
    \parbox{280pt}{%
\textbf{HM-240b} - The number of FCCs  with a failed OSAS component decreases monotonically.
}%
}
\end{center}
\label{req:mono}
\end{requirement}

The rationale behind Requirements~\hyperref[req:bound]{HM-240a}~and~\hyperref[req:mono]{HM-240b} is that they offer some guarantee about the controllability of the vehicle. They are also written in precise language that can be verified using AGREE.  We modeled Requirements~\hyperref[req:bound]{HM-240a}~and~\hyperref[req:mono]{HM-240b}  as Guarantees~\ref{guar:bound}~and~\ref{guar:mono}, respectively, in the contract of the FCS component.

\begin{equation}
\begin{split}
\mathbf{guarantee\colon}\: low \leq avg \wedge avg \leq high \\
avg = \frac{(act_1 + act_2 + act_3 + act_4)} {4}
\end{split}
\label{guar:bound}
\end{equation}

The variable $act_n$ in Guarantee~\ref{guar:bound} represents the $n$th signal sent to an actuator. Each of these signals comes from a set of four redundant signals.  The values of $low$ and $high$ are constant values that are determined for each actuator. This guarantee is repeated for each actuator in the FCC.

The variable $num\_valid\_acts$ in Guarantee~\ref{guar:mono} represents the total number of valid actuator signals.  Readers familiar with Lustre will recognize the $\mathbf{pre}$ function~\cite{lustre}. The $\mathbf{pre}$ function returns the value of its expression on the previous time step, in this case the previous value of $num\_valid\_acts$.  This guarantee is also repeated for each set of quad redundant actuator signals.

\begin{equation}
\begin{split}
\mathbf{guarantee\colon}\: num\_valid\_acts \leq \mathbf{pre} (num\_valid\_acts)
\end{split}
\label{guar:mono}
\end{equation}

\subsection{Flight Control Computer}

The FCS consists of four individual FCCs.  The composition of the guarantees of the four FCCs prove the guarantees of the FCS. One of the FCC requirements is shown in Requirement~\hyperref[req:fcc-bound]{FCC-S-150}.

\begin{requirement}
\begin{center}
\noindent\fbox{%
    \parbox{280pt}{%
\textbf{FCC-S-150} - The FCC OSAS application shall perform FCC command fault detection and mitigation logic. 
}%
}
\end{center}
\label{req:fcc-bound}
\end{requirement}

This statement also lacks the precision needed to develop a direct formalization.  It is not obvious what constitutes ``mitigation logic'' or what is considered ``fault detection''. This requirement needed to be linked to more precise definitions in lower level requirements.  The OSAS component requirements, which are discussed in the next subsection, contain language describing how the output signal gains are computed. When the OSAS is declared faulty, its actuator signals are latched to zero. When the OSAS is behaving correctly each output signal is multiplied by a factor determined by the number of faulty FCCs.

The requirements of each FCC act as lemmas about the ISAS, OSAS, and CLAW components to prove the top level properties about the FCS component.  Based on the requirements of the OSAS and Requirement~\hyperref[req:bound]{HM-240a} in the FCS, Requirement~\hyperref[req:fcc-bound]{FCC-S-150} was modeled by Guarantees~\ref{guar:fcc-bound} and~\ref{guar:fcc-mono}. These guarantees fulfill some of the ``mitigation logic'' and ``fault detection logic'' functionality mentioned in Requirement~\hyperref[req:fcc-bound]{FCC-S-150}. The composition of these guarantees from all four FCCs is strong enough to prove Requirements~\hyperref[req:bound]{HM-240a} and~\hyperref[req:mono]{HM-240b} in the FCS, and they are abstract enough to be proven by some of the OSAS requirements.

\begin{equation}
\begin{aligned}
&\mathbf{guarantee\colon} \\
&\:\:\:\:(num\_valid = 0 \rightarrow (low \leq act \wedge act \leq 4*high)) \wedge \\
&\:\:\:\:(num\_valid = 1 \rightarrow (low \leq act \wedge act \leq 2*high)) \wedge \\
&\:\:\:\:(num\_valid = 2 \rightarrow (low \leq act \wedge act \leq (3/4)*high)) \wedge \\
&\:\:\:\:(num\_valid = 3 \rightarrow (low \leq act \wedge act \leq high))
\end{aligned}
\label{guar:fcc-bound}
\end{equation}

\begin{equation}
\begin{split}
\mathbf{guarantee\colon}\: \mathbf{pre}(act\_fail) \rightarrow act\_fail
\end{split}
\label{guar:fcc-mono}
\end{equation}

Guarantees~\ref{guar:fcc-bound}~and~\ref{guar:fcc-mono} are repeated for each actuator.  The variable $num\_valid$ represents the number of valid actuator signals from other FCCs, $act$ represents the signal being sent to the actuator, $low$ represents the lower bound of the actuator signal, $high$ represents the upper bound of the actuator signal, and $act\_fail$ is a Boolean variable that is true if the actuator signal is latched failed.

\subsection{Output Signal Analysis and Selection}

Each actuator signal is computed by the OSAS component.  The redundant actuators apply force to their associated control surface in parallel. The requirements for the OSAS component determine the gain to be applied to each actuator signal depending on whether there have been failures in the other FCCs.  The OSAS component also has requirements that state how the value of an actuator signal is determined in the event of a failure in the OSAS component's own FCC. Requirements~\hyperref[req:osas-gain]{OSAS-S-180},~\hyperref[req:osas-fail]{OSAS-S-140}, and~\hyperref[req:osas-ccdl-fail]{OSAS-S-170} reflect some of the requirements used to determine the gain of each actuator signal. Their formalizations are shown as Guarantees~\ref{guar:osas-gain},~\ref{guar:osas-fail}, and~\ref{guar:osas-ccdl-fail}, respectively.

\begin{requirement}[h]
\begin{center}
\noindent\fbox{%
    \parbox{280pt}{%
\textbf{OSAS-S-180} - OSAS shall compute the actuator command gain as the ratio of the total number of command channels to the number of valid command channels (i.e. 4/(number of valid command channels)).
}%
}
\end{center}
\label{req:osas-gain}
\end{requirement}

\begin{requirement}[h]
\begin{center}
\noindent\fbox{%
    \parbox{280pt}{%
\textbf{OSAS-S-140} - When an actuator command has been latched failed, OSAS shall set that actuator command to 0 (zero).
}%
}
\end{center}
\label{req:osas-fail}
\end{requirement}

\begin{requirement}[!h]
\begin{center}
\noindent\fbox{%
    \parbox{280pt}{%
\textbf{OSAS-S-170} - If the local CCDL has failed, OSAS shall set the local actuator command gain to 1 (one).
}%
}
\end{center}
\label{req:osas-ccdl-fail}
\end{requirement}

\begin{equation}
\begin{aligned}
&\mathbf{guarantee\colon} \\
&\:\:\:\:(num\_valid = 0 \rightarrow fcc\_gain = 4) \wedge \\
&\:\:\:\:(num\_valid = 1 \rightarrow fcc\_gain = 2) \wedge \\
&\:\:\:\:(num\_valid = 2 \rightarrow fcc\_gain = 4/3) \wedge \\
&\:\:\:\:(num\_valid = 3 \rightarrow fcc\_gain = 1)
\end{aligned}
\label{guar:osas-gain}
\end{equation}

\begin{equation}
\begin{split}
\mathbf{guarantee\colon}\: (latched\_failed \rightarrow fcc\_gain = 0) 
\end{split}
\label{guar:osas-fail}
\end{equation}

\begin{equation}
\begin{split}
\mathbf{guarantee\colon}\: (ccdl\_failed \rightarrow fcc\_gain = 1) 
\end{split}
\label{guar:osas-ccdl-fail}
\end{equation}

There are other requirements that determine the true actuator gain value for the OSAS component, but they have been omitted here for the sake of space.  These guarantees are used to prove Guarantee~\ref{guar:fcc-bound} in the FCC, and the composition of the FCC contracts are used to prove Guarantee~\ref{guar:bound} in the FCS.

In the next section we discuss errors that were discovered through the process of formalizing and analysing the QFCS requirements in AGREE.

\section{Analysis Results}\label{sec:probs}

We ran our analysis on a laptop computer with an Intel\textsuperscript{\textregistered} i5 CPU and 16GB of RAM. The tool was run inside a virtual machine running Ubuntu Linux. Using JKind as the model checker and Yices~\cite{yices} as the SMT Solver, the contract for the FCS was proved in 7 seconds and the contract for the FCC (all four FCCs had identical contracts) was proven in 115 seconds. Kind 2.0 had similar performance. In Table~\ref{tbl:results} we list some information about the size of the QFCS AADL model and the number of requirements that we formalized for each component. The \textit{Inputs} and \textit{Outputs} columns list the number of input and output features, respectively, that are present in the AADL model. Many of these features are complex structures that consist of multiple data fields. For example, one actuator output consists of 20 real number values. The number of variables generated in the Lustre code that is sent to the model checker is on the order of hundreds for each component. The \textit{Guarantees} column reports the number of guarantees in each component contract. This number roughly corresponds to the number of English language requirements that we formalized for each component from the requirements hierarchy described in Figure~\ref{fig:reqs-hier}.  The number of guarantees is not exactly the same as the number of English language requirements because the language of some of the requirements was changed somewhat during formalization (as discussed in Section~\ref{sec:reqs}).

\begin{table}[H]
\vspace*{-5mm}
\center
\begin{tabular}{|c|c|c|c|}
\hline
\textbf{Component} & \textbf{Inputs}   & \textbf{Outputs} & \textbf{Guarantees} \\ \hline
FCS                & 13              & 12      & 2                         \\ \hline
FCC                & 11              & 22      & 9                           \\ \hline
OSAS               & 9               & 4       & 9                             \\ \hline
ISAS               & 9               & 18      & 11                           \\ \hline
CLAW               & 1               & 1       & 1                           \\ \hline
\end{tabular}
\vspace{+1em}
\caption{Information about the QFCS AADL Model}
\label{tbl:results}
\end{table}

Through the course of our analysis, we discovered a number of problems with the QFCS requirements.  These errors were discovered either through formalizing the requirements, attempting to prove properties, or using AGREE's realizability analysis (which we describe briefly later in this section). In this section we give a few examples of the kinds of problems that we found.

\subsection{Errors Found During Formalization}

Some requirements contained clear mistakes that were found through formalizing the English text.  In our experience, this is almost always a benefit of formalizing  requirements.  One of these requirements is shown in Requirement~\hyperref[req:form-prob]{ISAS-S-260}.

\begin{requirement}
\begin{center}
\noindent\fbox{%
    \parbox{345pt}{%
\textbf{ISAS-S-260} - ISAS shall determine the selected value for a quad digital signal using the following table:
\begin{enumerate}
\item 4 good values with total range less than $SignalTolerance$, average all 4
\item 4 good values with total range greater than $SignalTolerance$, average middle 2
\item 3 good values with total range less than $SignalTolerance$, average all 3
\item 3 good values with total range greater than $SignalTolerance$, select middle value
\item 2 good values with total range less than $SignalTolerance$, average values
\end{enumerate}
}%
}
\end{center}
\label{req:form-prob}
\end{requirement}

Interpreting this requirement at face value would indicate that the selected signal from a set of quad redundant digital signals would be completely unconstrained in the event that the range of all four values of the quad-redundant signals were exactly equal to $Signal Tolerance$. This does not seem to be the intent of the requirement as it is stated. This problem was discovered while formalizing the requirement, but it would have otherwise been discovered while verifying assumptions about the CLAW component input signal ranges.

\subsection{Errors Found During Model Checking}\label{subsec:model-check}

For some QFCS components we were able to check whether or not the implementation met its requirements.  In addition to the requirements for the ISAS component, we were also provided with an algorithmic specification (in tabular format) for its implementation.  We formalized this specification and attempted to prove that it met its formalized requirements. This analysis can be performed in AGREE by determining if a component's guarantees are entailed by \textit{assertions} about the component's implementation. In essence, component assertions are treated similarly to the component assumptions described in Section~\ref{sec:comp-v}, but are not checked to determine whether they hold as result of the system level assumptions. Unlike component assumptions, which must be proven to hold by Formula~\ref{fml:correct-assump}, component assertions are thought of as ``details about how a component is designed.''

The ISAS component is responsible for determining a selected sensor value to send to the CLAW component from a set of redundant input signals. Some input signals are quad redundant while others are dual redundant.  Among the quad redundant signals are values from the Embedded GPS/INS Sensor (EGI).  For each dual redundant signal, there exists a roughly equivalent signal that can be computed from the EGI.  In the event that the two values of a dual input signal miscompare (are not equal within some tolerance) the equivalent value of the EGI is selected to be sent to the CLAW component. During verification it was discovered that the implementation for the ISAS component did not correctly implement Requirement~\hyperref[req:model-prob]{ISAS-S-220}. The implementation for the ISAS component did not meet this requirement when the following scenario occurred.

\begin{itemize}
\item Channels 1 and 2 of a dual redundant signal are neither stale nor out-of-range
\item Channels 1 and 2 of a dual redundant signal miscompare
\item The equivalent value from the EGI is not declared faulty
\item Channel 1 of a dual redundant signal miscompares with its equivalent EGI parameter
\item Channel 2 of a dual redundant signal does not miscompare with its equivalent EGI parameter
\end{itemize}

\begin{requirement}
\begin{center}
\noindent\fbox{%
    \parbox{300pt}{%
\textbf{ISAS-S-220} - In the case of mismatched dual input signals, ISAS shall set the selected value equal to the equivalent selected value of EGI data.
}%
}
\end{center}
\label{req:model-prob}
\end{requirement}

In this scenario, the implementation selects the average of Channel 2 of a dual redundant signal with its equivalent EGI parameter. AGREE produced a counterexample showing this behavior. Through discussions with the domain experts, it was determined that the implementation was correct, and the requirement should be amended to handle this scenario in the same manner.

\subsection{Errors Found During Realizability Analysis}\label{subsec:realize}

AGREE also has an analysis option to determine if a component's contract is \textit{realizable}. This analysis is detailed in other work~\cite{gacek15}. Informally, a component's contract is realizable if there exists some implementation for the component that obeys the contract. Realizability is a stronger notion than consistency. For example, consider a component with a single integer input and a single integer output.  Suppose the component's contract guarantees that the output is always half the value of its input.  The component's contract is consistent because there are certainly some values for the input that satisfies this contract (e.g., if the input is 2 then the output would be 1).  However, if the input is an odd value then there is no corresponding integer value for the output. This contract is not realizable because there is no way to implement a component that could compute output values to satisfy this contract for every allowable input value.

A diligent reader may have noticed that Requirements~\hyperref[req:osas-fail]{OSAS-S-140} and~~\hyperref[req:osas-ccdl-fail]{OSAS-S-170} are stated, likewise Guarantees~\ref{guar:osas-fail} and~\ref{guar:osas-ccdl-fail} are formulated, in a way that makes them unrealizable.  What happens in the scenario where an actuator is latched failed and the CCDL fails?  There is a contradiction in what the selected gain value should be (should it be $0$ or $1$?). This error eluded the engineers who originally drafted the requirements as well as the engineers who formalized them.  However, AGREE's realizability analysis was able to identify the error and provide a counterexample with variables $latched\_failed$ and $ccdl\_failed$ set to true.

After discussing the error with the domain experts who wrote the requirements, it was determined that the solution was to set an order of precedence for how the gain value is computed.  For example, Guarantees ~\ref{guar:osas-gain},~\ref{guar:osas-fail} and~\ref{guar:osas-ccdl-fail} could be reformalized as Guarantees ~\ref{guar:prob-gain},~\ref{guar:prob-fail} and~\ref{guar:prob-ccdl-fail}.

\begin{equation}
\begin{aligned}
&\mathbf{guarantee\colon} \\
&\:\:\:\:not\: (latched\_failed \vee ccdl\_failed) \rightarrow\\
&\:\:\:\:\:\:\:\:(num\_valid = 0 \Rightarrow fcc\_gain = 4) \wedge \\
&\:\:\:\:\:\:\:\:(num\_valid = 1 \Rightarrow fcc\_gain = 2) \wedge \\
&\:\:\:\:\:\:\:\:(num\_valid = 2 \Rightarrow fcc\_gain = 4/3) \wedge \\
&\:\:\:\:\:\:\:\:(num\_valid = 3 \Rightarrow fcc\_gain = 1)
\end{aligned}
\label{guar:prob-gain}
\end{equation}

\begin{equation}
\begin{aligned}
&\mathbf{guarantee\colon} \\
&\:\:\:\:(latched\_failed \Rightarrow fcc\_gain = 0)
\end{aligned}
\label{guar:prob-fail}
\end{equation}

\begin{equation}
\begin{aligned}
&\mathbf{guarantee\colon} \\
&\:\:\:\:(not\: latched\_failed \wedge ccdl\_failed \Rightarrow fcc\_gain = 1)
\end{aligned}
\label{guar:prob-ccdl-fail}
\end{equation}

\section{Lessons Learned}

Through the course of this project we developed a number of insights about the challenges and benefits associated with formalizing and proving requirements compositionally.

\begin{itemize}
\item Many of the requirements that we attempted to formalize were not conducive to compositional verification.  Some of the high level requirements contained language that included details about lower level components. These types of requirements are hard to prove compositionally because they require details about components that are at a low level in the hierarchy to be exposed at a high level. Care should be taken when drafting requirements to make sure that they are precise but still abstract enough to reasoned about compositionally.

\item Often when we found that a component implementation failed to meet its requirements, the requirements were amended to be satisfied by the implementation. This scenario seemed to occur frequently because the requirements were not expressed formally in the first place. The examples given in Sections~\ref{subsec:model-check} and~\ref{subsec:realize} are illustrations of this.

\item Requirements are hard to formalize without a clear description of the model's architecture. We started this project without descriptions of the component interfaces. Upon receiving the interface descriptions, it was clear that many of our original formalizations were not correct.

\item Often times proof failures will expose errors in the model. For example an incorrect connection between two components will often cause the model checker to produce a counter example to properties that would normally seem trivial. Formalizing and proving requirements gives some assurance that the architectural model is correct.
\end{itemize}
\section{Conclusion}\label{sec:conclusion}

Much of the effort in this work was spent trying to find reasonable formalizations for the original English language properties. The formalization process itself identified significant problems with the requirements as they were originally stated. Even after formalization, model checking and realizability analysis identified a number of other issues.

Future work includes modeling more of the QFCS architecture in AADL, and formalizing other requirements in AGREE.  In this project all of the components were modeled to execute synchronously. Based on discussions with the QFCS designers this seemed to be a fair assumption. However, many systems are composed of components that execute on different clock domains. Support for modeling components that execute asynchronously (or quasi-synchronously~\cite{caspi01}) is currently being added to AGREE.

\subsubsection{Acknowledgement.}
This work was sponsored by NASA under contract NNA13AA21C.
The authors are especially thankful to Robert Antoniewicz at NASA Armstrong Flight Research Center for many helpful discussions about the QFCS design.

\bibliography{document}
\bibliographystyle{splncs}

\end{document}